\documentclass[A4,10pt]{article}
\usepackage[dvips]{graphicx}
\input epsf.sty 

\newcommand{\be}{\begin{eqnarray}}
\newcommand{\ee}{\end{eqnarray}}
\newcommand{\nn}{\nonumber\\}

\newcommand{\la}{\langle}
\newcommand{\ra}{\rangle}

\begin{document}

\begin{center}

{\LARGE {\bf Random Z(2) Higgs Lattice Gauge Theory \\
in Three Dimensions and its Phase Structure
}}
\vspace{1cm}

{\Large Shunsuke Doi, Ryosuke Hamano, Teppei Kakisako,\\
Keiko Takada, and Tetsuo Matsui} 

\vspace{1cm}

Department of Physics, Kinki University, 
Higashi-Osaka, 577-8502 Japan 

\vspace{1cm}


\end{center}


\begin{abstract}
We study the three-dimensional random Z(2) lattice gauge 
theory with Higgs field, which has the link Higgs coupling
$c_1 SUS$ and the plaquette gauge coupling $c_2 UUUU$.
The randomness is introduced by replacing
$c_1 \rightarrow -c_1$ for each link with the probability $p_1$
and  $c_2 \rightarrow -c_2$ for each plaquette 
with the probability $p_2$. We calculate the phase diagram by 
a new kind of mean field theory that does not assume the replica symmetry
and also by Monte Carlo simulations.   
For the case  $p_1=p_2(\equiv p)$, the Monte  Carlo simulations exhibit that
(i) the region of the Higgs phase in the Coulomb-Higgs
transition diminishes  as $p$ increases, and
(ii) the first-order phase transition between the Higgs 
and the confinement phases disappear for $p \ge p_c \simeq 0.01$.
We discuss the implications of the results to the 
quantum memory studied
by Kitaev et al. and the Z(2) gauge neural network on a lattice.
\end{abstract}


\eject

\section{Introduction}

Randomness and/or disorders are involved in quite many physical systems and 
play an important role in various fields of physics.
In some electron systems, randomness appears due to
impurities and affects electric conductivity.
Sufficient randomness 
 may cause localization of electrons\cite{anderson}. 

For a quantum computer, the effects of environmental
noises (including
thermal fluctuations, disorders by impurities, etc.), which
let the system to decohere, should be reduced to  
perform quantum computation as one designed. 
Kitaev\cite{kitaev} proposed
a fault-tolerant quantum memory and quantum computations that are 
based on the Aharonov-Bohm effect of discrete $Z_2$ gauge symmtery.
The memory is on a two dimensional torus and   modeled as the
Z(2) lattice gauge theory on a three dimensional (3D) lattice
with the third dimension for the imaginary time.
It is argued that the quantum memory works well
if this  model  system is in the Coulomb phase\footnote{The ordered phase
discussed in the Z(2) model of Ref.\cite{kitaev}
is the Coulomb phase instead of the Higgs phase as we shall clarify later.
Because of the discreteness
of Z(2) group, the mass of gauge boson becomes finite even in the 
Coulomb phase. }
(an oredered phase with small  fluctuations of gauge-field)
instead of the confinement phase (a disordered phase with
large fluctuations of gauge field).  
After that, many studies on Kitaev's model 
have appeared\cite{preskill}.
Wang et al.\cite{wang} studied accuracy threshould of the model
by using the $Z_2$ random-plaquette 
gauge model(RPGM) on a three-dimensional lattice.
In the Z(2) RPGM, the (inverse) gauge coupling for each plaquette
takes the values $\pm\beta$ with
random sign; the probability to take $\beta$ is given by $1-p$ 
and the probability of  ``wrong-sign" $-\beta$ is $p$.
The main interest is its phase structure, i.e.,
the critical concentration $p_c(\beta)$ which disinguishes
the confinement phase and the Coulomb  phase. 
Wang et al.\cite{wang} calculated $p_c(\beta=\infty) \simeq
0.029$. Ohno et al.\cite{ohno} calculated $p_{c}(\beta)$
for finite $\beta$ and showed
$p_c(\beta)$ takes a maximum value $p_c(\beta) \simeq 0.032\sim 0.033$ 
at $\beta \simeq 0.4\sim 0.5$.
(We shall explain this $p_c(\beta)$ in detail
as the special case $c_1=0$ of the present model
in Sect.4 using Fig.\ref{fig4}.)

In this paper we consider another  random gauge model
related to the Z(2) RPGM, 
the 3D random Z(2) Higgs lattice gauge theory. 
In addition to the usual gauge coupling
on each plaquette (i.e., the coupling of four gauge fields on 
the links around each plaquette) as the Z(2) RPGM, 
the energy of this model 
contains the Higgs coupling
on each link (a pair of Higgs fields at the nearest-neighbor
sites couples through the gauge field on the connecting link).
Both the coupling constants of Higgs and gauge couplings
have wrong signs with certain probabilities. 
The pure case of the model is known to have three phases,
Higgs, confinement, and the Coulomb phase\cite{kemukemu}.
We study its phase structure, in particular, how the phase boundaries
shift as the randomness is increased. 

The reasons why we are interested in this model are as follows:
First reason is  related to the quantum memory.
Are there any real materials that can be a candidate
of Kitaev's model? 
Many strongly correlated electron systems can be 
described by an effective gauge field theory\cite{nagaosa}.
For example, the low-energy behavior of 
$s=1/2$ antiferromagnetic  Heisenberg spin model
is described effectively  by the U(1) lattice gauge theory
coupled with CP$^1$ spins.\cite{takashima}
However, most of such low-energy effective gauge field theories
contains gauge coupling of matter fields in addition to
the pure gauge term.
When the gauge group is Z(2), the corresponding
lattice gauge theory is the Z(2) Higgs lattice gauge theory,
which is just the model  we are going to 
investigate.
In  Kitaev's model, the Wilson loops are used as the index of 
quantum memory as well as the order parameter of confinement-deconfinement
phase transition.
However, for the gauge theory including matter couplings,
it is well known that the Wilson loop    always obeys
the perimeter law and cannot be used as an order parameter.
Another nonlocal order parameters are proposed\cite{marc}.
We expect that this Higgs  system is more realistic and 
can work as a quantum memory
when it is realized in the Higgs phase.

The second reason is related to  neural network models of the 
human brain.\cite{nn}
The Hopfield model\cite{hopfield} is well known as the standard neural network
 model to explain
the mechanism of associative memory, i.e., 
a process of retrieving a pattern of neurons once stored in the brain.
For the process of learning a pattern itself, 
various models based on the plasticity
of synaptic connections have been proposed.
In Refs.\cite{kemukemu,nnet}, we introduced and studied a neural network
with gauge symmetry, in which the synaptic variables are treated as
a gauge field. This sounds natural because electromagnetic
signals propagate through the  synaptic connections, and the 
electromagnetic interaction has U(1) gauge symmetry.
There is a strong correlation between the performance of 
learning a pattern and retrieving it and the three phases of gauge dynamics.

In neural networks, there should be necessarily
malfunctions of signal propagations.
In Ref.\cite{nnet}, they  
are described as thermal fluctuations at finite pseudo temperatures
by employing the Metropolis algorithm as the rule of time evolution
as in the Boltzmann machine.
Another possibility of description of malfunctions 
may be introducing impurities
in the network. When the network is put on a lattice
and the synaptic connections are restricted to  the 
nearest-neighbor ones, such a model
becomes just the random  Z(2) Higgs lattice gauge theory we are going
to study.  
It is interesting to see how the randomness caused by impurities
affects the functions of human brain.

The present paper is organized as follows.
In Sec.2, we introduce the model.
In Sec.3, we set up a mean field theory for systems with randomness
and applies it to the present model.
As the mean-field theory,
the replica method for quenched averaged systems like spin glass
is well known. The present mean-field theory is quite different 
from the replica method. It is general and applicable for any random
system having ``wrong-sign" coupling constants
as long as usual mean field theory for the pure sustem is available.
In Sec.4, we study the phase diagram by Monte Carlo simulations.
In Sec.5, we preent discussion and conclusions.

\section{The Model}
\setcounter{equation}{0} 

The model is defined on the three-dimensional cubic lattice
of the size $V = L \times L \times L$
with the periodic boundary condition.
On each site $x$ there sits a Z(2) spin variable $S_x = \pm 1$
and on each link $(x,x+\mu)$ ($\mu = 1,2,3 $ is the direction index
and we use it also as the unit vector) there sits a Z(2) gauge
variable $U_{x\mu} = \pm 1$. 
The energy (multiplied by minus of the inverse effective temperature) 
$A$ of the model is given by
\be
A=\sum_{x}\sum_{\mu=1}^3 c_{1x\mu} S_{x+\mu} U_{x\mu} S_x
+\sum_{x}\sum_{\mu < \nu}c_{2x\mu\nu}U_{x\nu}U_{x+\nu,\mu}
U_{x+\mu,\nu}U_{x\mu},
\label{action}
\ee
where $c_{1x\mu}$ and $c_{2x\mu\nu}$ are random coefficients
on each undirected link $(x,x+\mu)$ and 
unoriented plaquette $(x,x+\mu,x+\mu+\nu,x+\nu,x) (\mu < \nu)$, 
respectively.
 The energy $A$ of (\ref{action}) is invariant under the 
local gauge transformation,
\be
S_x \rightarrow S'_x = W_x S_x,\ 
U_{x\mu} \rightarrow U'_{x\mu} = W_{x+\mu} U_{x\mu} W_x,\ 
W_x =\pm 1. 
\ee
$c_{1x\mu}$ and $c_{2x\mu\nu}$
are independent random variables taking the values as
\be
c_{1x\mu}&=& \left\{
\begin{array}{ll}
c_1\ &{\rm with\ the\ probability}\ 1- p_1 \\
-c_1\ &{\rm with\ the\ probability}\ p_1 
\end{array}
\right.\nn
c_{2x\mu\nu}&=& \left\{
\begin{array}{ll}
c_2\ &{\rm with\ the\ probability}\ 1 - p_2 \\
-c_2\ &{\rm with\ the\ probability}\ p_2 
\end{array}
\right.
\label{c1c2}
\ee
We regard the link with the ``wrong  sign", $c_{1x\mu} = -c_1$,
 a link with  impurity, and similarly,
the plaquette with $c_{2x\mu\nu} = -c_2$ a plaquette 
with impurity.
Therefore $p_1$ and $p_2$ are the concentrations of link and 
plaquette impurities, respectively.   
$c_1$ and $c_2$ are positive parameters appearing
in the pure system ($p_1=p_2=0$).
We noite that, if one sets $c_1=0$ then the present model
reduces to the Z(2) random plaquette model considered
in Refs.\cite{kitaev,wang}.
Each sample lattice has a fixed  configuration of $c_{1x\mu}$
and $c_{2x\mu\nu}$. To calculate a physical quantity
like the internal energy $E$ and the specific heat $C$
which are measured in experiments,
we first consider the thermal(annealed) 
average $\la O(S,U) \ra$ of observable $O(S,U)$ for each sample as
\be
\la O(S,U) \ra_i &\equiv 
&\frac{1}{Z_i}\sum_S \sum_U O(S,U) \exp(A_i),\nn
Z_i &\equiv&\sum_S \sum_U  \exp(A_i) \equiv \exp(-F_i),
\ee
where $i$ is the suffix to specify the sample as the $i$-th sample,
and $Z_i$ is its partioton function and $F_i$ is its free energy.
Then we take the sample average $\overline{\la O(S,U) \ra}$
of $\la O(S,U) \ra_i$, 
a quenched average over sufficiently larage number $N_S$ of samples, 
\be
\overline{\la O(S,U) \ra} \equiv 
\lim_{N_S \rightarrow \infty}\frac{1}{N_S} \sum_{i=1}^{N_S}
\la O(S,J) \ra_i.
\label{quench}
\ee
This expression corresponds to  physically measured 
quantities in a random system. The sample average (\ref{quench})
can be rewritten as
\be
\overline{\la O(S,U) \ra}  
= \int [dc_1][dc_2]\rho_{12}(c)\la O(S,J) 
\ra(c),
\ee
where we wrote $\la O(S,J) \ra(c) 
\equiv \la O(S,J) \ra_i$,
$[dc_1][dc_2] \equiv \prod_{x,\mu} dc_{1x\mu}
\prod_{x,\mu < \nu} dc_{2x\mu\nu},$
and introduced the probability
distribution of $c_{1x\mu}$ and $c_{2x\mu\nu}$,
\be
\rho_{12}(c) = \prod_{x,\mu}[(1-p_1)\delta(c_{1x\mu}-c_1)
+p_1\delta(c_{1x\mu}+c_1)]
\prod_{x,\mu < \nu}[(1-p_2)\delta(c_{2x\mu\nu}-c_2)
+p_2\delta(c_{2x\mu\nu}+c_2)].
\ee
The free energy $F$, the internal energy $E$ per site, and 
the specific heat $C$ per site are defined by
\be
F = \overline{F(c)},\ 
E = -\overline{\la A \ra(c)}/V,\  
C = \overline{\la (A - \la A \ra)^2 \ra(c)}/V. 
\ee

\section{Mean Field Theory for a Random System with Wrong-Sign Coupling
Constants}
\setcounter{equation}{0} 

Usually, the mean field theory (MFT) applied for a statistical system
gives a rough but intuitive
understanding of the global properties of the system
like its phase structure.
We first summerize the MFT for a pure system and then
develope a new MFT for a random system based on that for a pure system.

\subsection{MFT for a pure system}

For pure models without  impurities
 a mean field theory can be formulated
by the variational principle\cite{feynman}.
Let us briefly summerize it.
We start with the partition function $Z$
for the action $A(\phi)$ with a set of  variables $\phi$, 
\be
Z = \sum_{\phi} \exp(A(\phi)) \equiv \exp(-F),
\ee
where $F$ is the free energy.
In MFT one prepares a trial action $A_0(\phi, \lambda)$
having (a set of) variaitonal parameters $\lambda$, the 
partition function $Z_0$ of which is calculable; 
\be
Z_0 = \sum_{\phi} \exp(A_0(\phi,\lambda)) \equiv \exp(-F_0(\lambda)).
\ee
Then there holds the following Jensen-Peiels inequality; 
\be
F &\leq& F_v(\lambda) \equiv F_0(\lambda) + \la A_0(\lambda)-A \ra_0,
\nn
\la O(\phi) \ra_0 &\equiv & \frac{1}{Z_0} \sum_{\phi} O(\phi)
\exp(A_0(\phi,\lambda)).
\label{ineq}
\ee
Thus, one minimizes $F_v(\lambda)$ by adjusting $\lambda$
to obtain the best approximation for $F$.

For the present model (\ref{action}) at $p_1 = p_2 =0$,
it has been  applied in Ref.\cite{kemukemu}
with the choice
\be
A_0 = a_0\sum_x S_x +a_1\sum_{x}\sum_{\mu} U_{x\mu},
\ee
where $a_0,\ a_1$ are the two variational parameters.
Explicitly, we have
\be
F_v/N &=& -\log[2\cosh(a_0)]-d\log[2\cosh(a_1)]
+a_0 \tanh(a_0)
+d a_1 \tanh(a_1)\nn
&&-d c_1\tanh^2(a_0)\tanh(a_1)
-\frac{d(d-1)}{2} c_2\tanh^4(a_1). 
\ee
The stationary conditions read
\be
a_0 &=& 2dc_1\tanh(a_0)\tanh(a_1),\nn
a_1&=&dc_1\tanh^2(a_0)+2d(d-1)c_2\tanh^3(a_1).
\label{minimizepure}
\ee
It predicts the three phases as listed in 
Table 1.\footnote{Some cases of Table 1 show
the averages of gauge-variant quantities are nonvanishing, 
which violate Elitzur's theorem\cite{elitzur}.
However, an additional averaging procedure over gauge rotated copies
of the result of MFT gives rise to vanishing averages without modifying the 
phase structure of MFT\cite{drouffe}.}
In Fig.\ref{fig1}
we plot the phase boundaries determined by MFT
in the $c_2-c_1$ plane. 

\begin{center}
 \begin{tabular}{|c|c|c|c|} 
\hline
   phase    & $\langle U_{x\mu} \rangle_0 $ & $\langle S_x \rangle_0 $  
   &  ability 
\\ \hline
Higgs       & $\neq 0$  & $\neq 0$  & learning and recalling   
\\ \hline
Coulomb     & $ \neq 0$  & $0$  & learning  \\ \hline
Confinement & $0$   & $0$     &  N.A.
\\
\hline
\end{tabular}\\
\end{center}
\vspace{0.5cm}
Table1. Three phases predicted in MFT and the associated ability
of neural net in a process of learning a pattern of $S_x$
and retrieving it\cite{nnet}.

\subsection{MFT for a random system}

For the case of random system that involves 
quenched averages like (\ref{quench}), one should generalize
the MFT  (\ref{ineq}) for a pure system. 
Below we develop such a generalization.
For this purpose we start with the following variational system,
which is independent of differences among samples;
\be 
Z_{0}(a) &=& \sum_{S_x} \sum_{U_{x\mu}} \exp[A_{0}(S,U,a)]
\equiv \exp[-F_{0}(a)],\nn
A_{0}(S,U,a)&=& \sum_x a_{0x} S_x +\sum_{x}\sum_{\mu} a_{1x\mu}U_{x\mu},
\label{zrandom}
\ee
where $a_{0x}$ and $a_{1x\mu}$ are {\it local} 
variational parameters  on the
site $x$ and link $(x,x+\mu)$ respectively.
(We assign the suffices 0, 1, and 2 for site,  link, and plaquette
 objects respectively.)
By applying the inequality (\ref{ineq}), one has
\be
F_i(c) \leq F_{v}(a,c) 
\equiv F_{0}(a)
+\la A_0\ra_0(a) -\la A \ra_{0}(a,c).
\ee
By multiplying $\rho_{12}(c)$ both sides and integrating over
$c_{1x\mu},c_{2x\mu\nu}$ one obtains the inequality
for the free energy $F$,%
\be
F \leq \int [dc_1][dc_2]\rho_{12}(c)F_{v}(a,c).
\label{frandom}
\ee
Below we  treat $a_{0x}$ and $a_{1x\mu}$ as random variables
described by the probability
distribution $\rho_{01}(a)$  defined as
\be
\rho_{01}(a) = \prod_{x}[(1-q_0)\delta(a_{0x}-a_0)
+q_0\delta(a_{0x}+a_0)]
\prod_{x,\mu}[(1-q_1)\delta(a_{1x\mu}-a_1)
+q_1\delta(a_{1x\mu}+a_1)],
\ee
We consider the sample average  of a function
$f(a)$ over samples of variational systems, each of which 
has different
set of $a_{0x}, a_{1x\mu}$, and write it as  $\widetilde{f(a)}$,
\be
\widetilde{f(a)} \equiv \int[da_0][da_1]\rho_{01}(a)f(a).
\ee  
By multiplying $\rho_{01}(a)$ both sides of (\ref{frandom}) 
and integrating over $a_{0x}, a_{1x\mu}$, one has
an upperbound $\widetilde{\overline{F}}_v$ for the free energy
$F$ as
\be
F \leq \widetilde{\overline{F}}_v \equiv \int [da_0][da_1][dc_1][dc_2]
\rho_{01}(a)\rho_{12}(c)F_{v}(a,c).
\label{var}
\ee
To relate $p_1,p_2$ and $q_0,q_1$ we regard
$q_0,q_1$ as concentrations of effective impurities
on the sites and links respectively, and 
compose real impurities as the products of them.
Explicitly, we determine the two functions
$q_0(p_1,p_2)$ and $q_1(p_1,p_2)$ so that the 
following relations hold in average over samples
with $\rho_{01}(a)$;
\be
(1)&{\rm average\ number\ of\ links\ 
for\ which\ sign}[a_{0x}a_{1x\mu}a_{0,x+\mu}]=-1\ {\rm is}\ 3Vp_1
\nn 
(2)&{\rm average\ number\ of\ plaquettes\ 
for\ which\ sign}[a_{1x\nu}
a_{1,x+\nu,\mu}a_{1,x+\mu,\nu}a_{1x\mu}]=-1\ {\rm is}\ 3Vp_2.
\label{qofp}
\ee
In Fig.\ref{fig2} we present $q_{1,2}(p_1,p_2)$ calculated by
using a 3D lattice of the size $24^3$.

Then $\widetilde{\overline{F}}_v$ of (\ref{var}) is calculable because
$\int [da_0][da_1]$ and $\int [dc_0][dc_1]$  are
straightforward due to the decoupled nature
of $F_0$, $A_0$ and $A$. 
\be
\widetilde{\overline{F}}_v/V &=& -\log[2\cosh(a_0)]-d\log[2\cosh(a_1)]
+a_0 \tanh(a_0)
+d a_1 \tanh(a_1)\nn
&&-d(1-2p_0)(1-2q_0)^2(1-2q_1)c_1\tanh^2(a_0)\tanh(a_1)\nn
&&-\frac{d(d-1)}{2}(1-2p_1)(1-2q_1)^4c_2\tanh^4(a_1). 
\ee
$\widetilde{\overline{F}}_v$ becomes 
a function of two parameters $a_0, a_1$, which
we adjust to minimize it. It is straightforward to 
see that the conditions of minimization 
take the same form as  the conditions (\ref{minimizepure}) 
 for the pure system ($p_1=p_2=0$);
the former is given from the latter (the pure system) with the replacements
\be
c_{1}&\rightarrow&c_1'\equiv G_1(p_1,p_2) c_{1},\nn
 c_{2} &\rightarrow& c_2'\equiv G_2(p_1,p_2)  c_{2},\nn
G_1(p_1,p_2) &\equiv& (1-2p_{1})[1-2q_{0}(p_1,p_2)]^{2}[1-2q_{1}(p_1,p_2)],
\nn
G_2(p_1,p_2)&\equiv& (1-2p_{2})[1-2q_{1}(p_1,p_2)]^{4}
\ee
Thus the phase transition points $(c_1^{\rm random}, 
c_2^{\rm random})$  of the random system ($p_1, p_2 \neq 0$)
are obatined by using the transitoin points $(c_{1}^{\rm pure},
c_{2}^{\rm pure})$
of the pure system  ($p_1=p_2=0$) as
\be
c_{1}^{\rm random}(p_1,p_2)&=&\frac{1}{G_1(p_1,p_2)} c_{1}^{\rm pure}\nn
c_{2}^{\rm random}(p_1,p_2)&=&
\frac{1}{G_2(p_1,p_2)} c_{2}^{\rm pure}.
\label{mft}
\ee
Because $G_{1,2}(p_1,p_2) \geq 1$, one obtains
the following general inequalities 
\be 
c_{1}^{\rm random}(p_1,p_2) \geq c_{1}^{\rm pure},\ \
c_{2}^{\rm random}(p_1,p_2) \geq c_{2}^{\rm pure}.
\ee
This implies that the effect of impurities/disorder
increases the critical coupling constants,
which accords with our intuition.
In Fig.\ref{fig3} 
we plot the critical lines of the present model 
in the $c_2-c_1$ plane
obtained by (\ref{mft})  for the cases of $p\ (\equiv p_1=p_2)=0.015 \sim 0.050$
together with the critical lines for $p=0$.

We note that the present MFT can be applied for a random 
lattice system similar to the present one as long as its
variational free energy is decoupled to single-site and  single-link
integrals like (\ref{zrandom}). In particluar, 
the probability distributions $q_{0,1}(p_1,p_2)$  
of effective impurities are universal, i.e.,
independent of each model, and determined by (\ref{qofp}).
Therefore, for each critical point 
of the pure system,  there  may be one associated
critial point of the random system
as a modulation due to randomness like (\ref{c1c2}).

\section{Monte Carlo Simulations}
\setcounter{equation}{0} 

In this section, we start with 
summerize the  known results for $p_1=p_2=0$ and for $c_1=0$.
Then we study the phase struture of the 3D Z(2)
random Higgs lattice gauge theory by MC simulations
for the case $p_1=p_2\equiv p$.

\subsection{The case of $p=0$}

Let us first summerize the phase structure for the pure case $p=0$.
The points in Fig.\ref{fig1} is the phase boundary 
in the $c_2-c_1$ plane
\cite{kemukemu}, exhibiting three phases.

The confinement-Higgs transition is first-order and terminates
near $(c_2,c_1)\simeq (0.55,0.35)$ as the complementarity argument 
predicts.\cite{complementarity}
In fact, at $c_2=0$, the partition function is exactly calculable
by the single-link sum over $U_{x\mu}$ as
\be
Z|_{c_2=0} = [2\cosh(c_1)]^{3N},
\label{tricritical}
\ee
which has no singularity
in the $c_1$-dependence and so there is no phase transition along $c_2=0$.   
The Coulomb-Higgs transition
is second order. In the limit $c_2 \rightarrow \infty$,
$U_{x\mu}$ becomes a pure-gauge configuration,
$U_{x\mu} = W_{x+\mu}W_x \ (W_x = \pm1)$, the system reduces to the 3D
Ising spin model $A=c_1\sum_{x,\mu} S'_{x+\mu}S'_x\ (S'_x=W_x S_x)$, 
which exhibit a second-order phase transition at $c_1 \simeq 0.22$. 
The confinement-Coulomb transition is also second order.
Along $c_1=0$, the system reduces to the pure gauge model, which
is known to exhibit a second-order phase transition at
$c_2 \simeq 0.76$.\\

\subsection{The case of $c_1=0$}

Next, let us summerize the case of no Higgs coupling ($c_1=0$)
but with randomness. In Fig.\ref{fig4} we present 
the phase diagram in the $p-T$ ($p\equiv p_2,\ T\equiv c_2^{-1}$) plane 
obtained in Ref.\cite{ohno}. 
At $p=0$, a second-order phase transition at 
the critical point
$T=T_c \simeq 1/0.76$.  
For $T < T_c$ the gauge-field fluctuations are small
and the system is in the ordered Coulomb phase, while 
for $T_c <  T$, the fluctuations are large
and the system is in the disordered confinement phase.
As $p$ increases from $p=0$, 
the critical coupling constant $T_c(p)$ 
decreases due to the randomness in gauge couplings, 
and it shall vanishes at a certain value $p_{0}$, $T_c(p_{0})=0$. 
The data of specific heat suggests
that the order of transition changes from second order to
higher order for $T$ smaller than about 1.15. 
$T_c(p)$ seems to show a ``reentrance behavior", i.e.,
as  $T$ is lowered, the value of $p$ on the curve  $T_c(p)$ 
increases first and takes the 
maximum value $p_{\rm max} \simeq 0.033$\cite{ohno}
around $T \simeq 0.4 \sim 0.5$ and then decreases
to end up with $p \simeq 0.029$ at $T=0$\cite{wang}.

\subsection{Set up of MC simulations and the global phase structure}

In our MC simulations we use the standard Metropolis 
algorithm\cite{metropolis}.
The typical number of sweeps in a single run for
$\la O \ra_i$ of each sample is 
$\sim 2 \times 10^5$. To  estimate 
the  error of  $\la O \ra_i$, which we call the MC error,
we divide a run into 10 successive intervals
to generate 10 data. AS the MC error we estimate 
 the standard deviation of these 10 data. 
 The average acceptance ratios are
$0.4 \sim 0.5$. 
In Fig.\ref{fig5}a we present a typical result of the specific heat
$C$ together with the MC errors. 
For a random system one may be interested in
the standarde deviation over samples (SDS), which
should converge to a nonvanishing value even in the limit $N_s 
\rightarrow \infty$.
In Fig.\ref{fig5}b we present such SDS for the same $C$ as Fig.\ref{fig5}a
with the sample number $N_s = 40$.
By comparing these two figures, 
both quantities are similar in magnitude
(they differ by up to factor $\sim 2$),
and  have similar behaviors
 (i.e., the $c_1$ dependence).
In the figures below the error bars show SDS.

To see the dependence of $\bar{\la O \ra}$ 
on $N_s$ and to find a suitable value of $N_s$
we present in Fig.\ref{fig6} the specific heat $C$ vs $N_s$.
It shows that the  $N_S$ dependence is rather weak
for $N_S \ge 20 $. Below we show the results for $N_s =40$
otherwise stated, where  the error bars in the figures show SDS.

Let us first observe how the peak 
of the specific heat shifts as $p$ increases.
In Fig.\ref{fig7}a 
we present the curves of the peak location of specific heat
in the $c_2-c_1$ plane. 
In Fig.\ref{fig7}b, we present the specific
heat for various $p$ at $c_2=1.0$.
As $p$ increases, 
its peak becoms rounder and its position shifts
to larger $c_1$ direction.
It is consistent with the MFT prediction of Fig.\ref{fig3} 
in Sect.3. 
In Fig.\ref{fig7}c, we present the specific
heat for various $p$ at $c_1=0.1$.
Its peak becomes rounder but its position
is almost unchanged. We shall discuss the new behavior of $C$
appearing in  larger latice systems.

\subsection{Study of the three cases}

Let us see whether these specific-heat peaks 
exhibit genuine phase transitions
or not. To be specific, we focus on the following three cases:\\
(A) The confinement-Higgs transition along $c_2=0.7$,\\
(B) The Coulomb-Higgs transition along $c_2=1.0$,\\
(C) The confinement-Coulomb transition along $c_1 =0.1$.\\

(A) $c_2=0.7$:

For $p=0$ we know that this case gives rise to a first-order
phase transition.
In Fig.\ref{fig8}, we present $U$ and $C$ for $p=0.01$. 
$U$ shows no hysteresis and
the peak of $C$ shows no systematic system-size dependence, which
imply that this peak implies a higher-order transition 
or a crossover but not a first-order
nor second-order phase transition. The small rendomness of $p=0.01$ 
is sufficient to destroy the first-order transtion of $p=0$ here.
Because the firsr-order transition line for $p=0$ ends at $c_2 \simeq 0.55$ 
and no phase transitions follow in the smaller $c_2$ region 
(see eq.(\ref{tricritical}))
it is quite possible that there are
no genuine phase transitions of finite order in  
the case (A) for $p \geq 0.01$. \\

\begin{figure}[b]
\includegraphics[width=9cm]{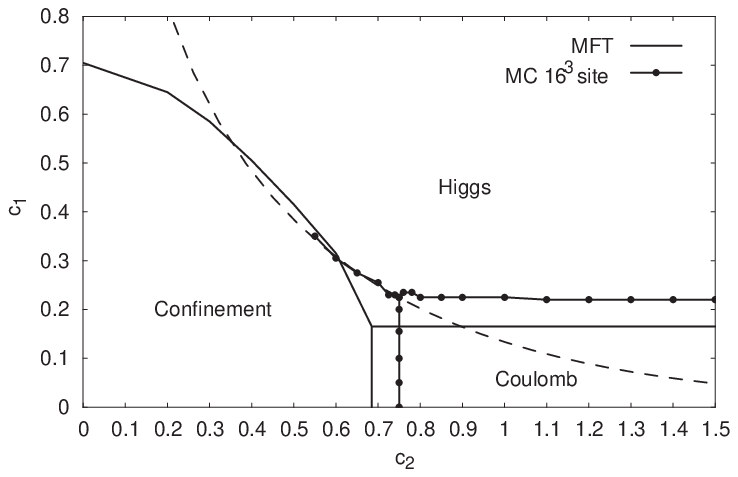}
\caption{\label{fig1}
Phase diagram by MFT for $p=0$. The dots represents the 
results of MC simulations.\cite{kemukemu} 
The dashed curve $c_2 = -\frac{1}{2}{\rm \ell n th}(c_1) $ is the exact
results obtianed by the self-duality argument\cite{duality}, 
on which some
of the transition points may lie.
}\end{figure}

\begin{figure}[h]
\includegraphics[width=6cm]{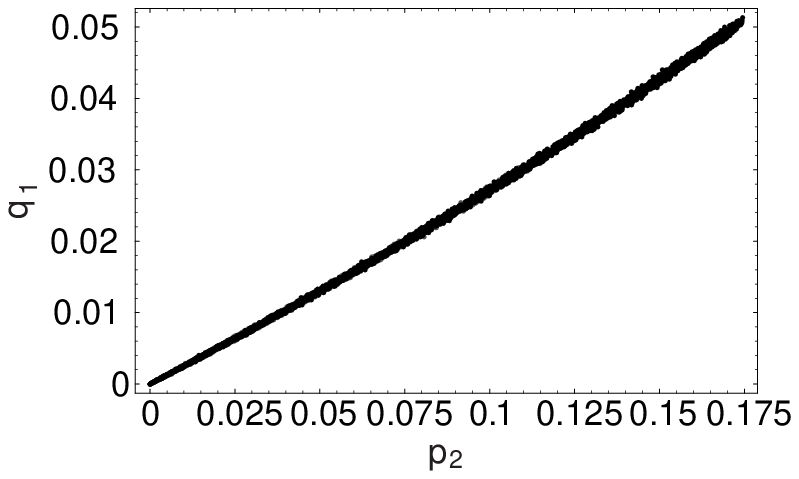}
\hspace{-1.5cm}
\hspace{1.2cm}
\includegraphics[width=6.5cm]{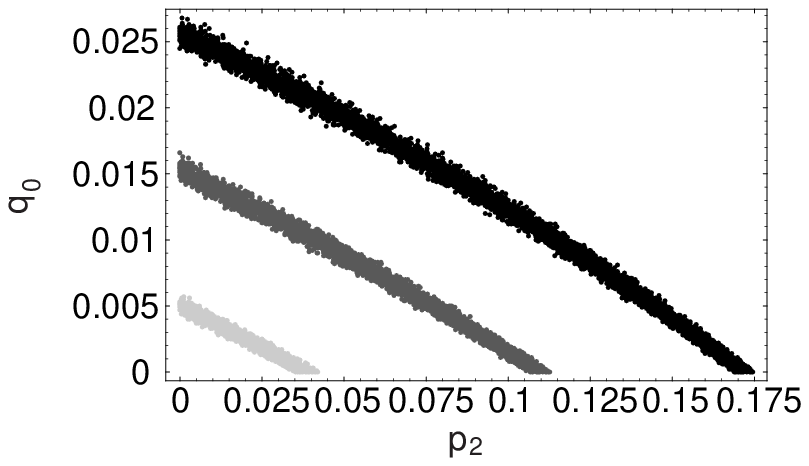}
\vspace{-0.5cm}\\
\hspace{3cm}(a) \hspace{5cm} (b) \\
\includegraphics[width=6cm]{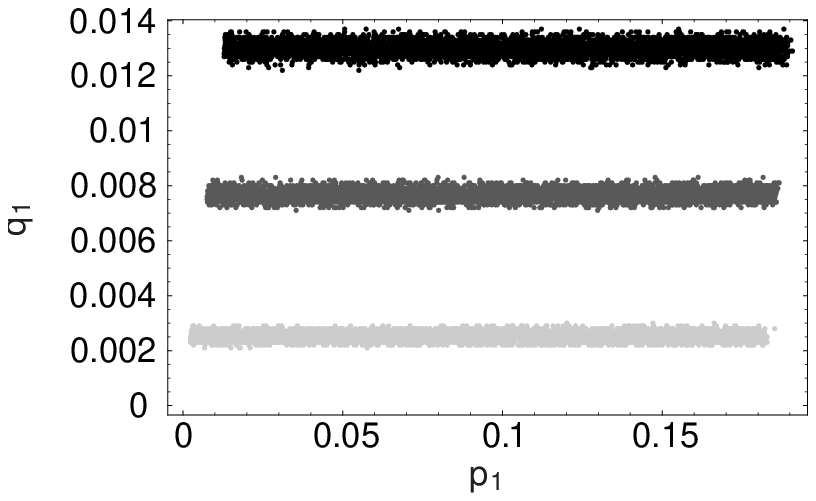}
\includegraphics[width=6cm]{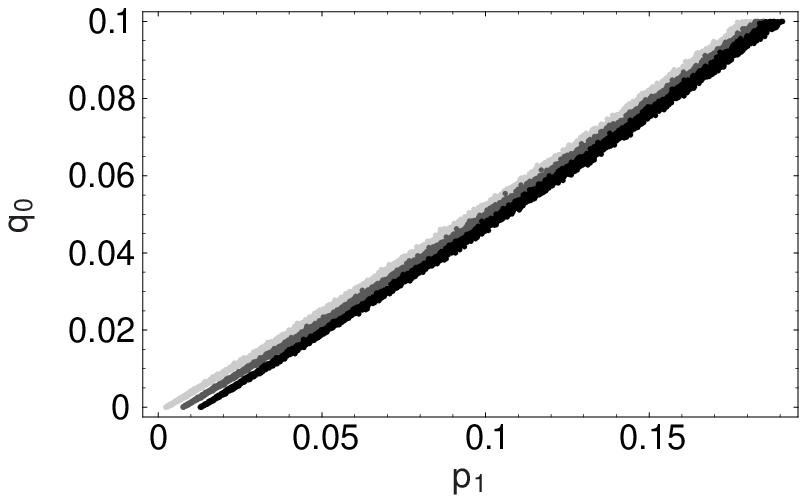}\vspace{-0.5cm}\\
\hspace{3cm}(c) \hspace{5cm} (d) \\
\vspace{-0.5cm}
\caption{\label{fig2}
$q_{0}(p_1,p_2)$ and $q_{1}(p_1,p_2)$ satisfying (\ref{qofp}) 
calculated by
using a 3D lattice of the size $24^3$.
(a) $p_2$ vs $q_1$ for $p_1=0.01,0.03,0.05$.
$q_1$ is independent of $p_1$ because
the wrong $p_2$-plaquettes are generated by $q_1$-links alone
without $q_0$-sites.
(b) $p_2$ vs $q_0$ for $p_1=0.01,0.03,0.05$ from below.
$q_0$ decreases as $p_2$ increases  because
too sufficient number of $q_1$-links are generated to make $p_2$-plaquettes.
(c) $p_1$ vs $q_1$ for $p_2=0.01,0.03,0.05$ from below.
$q_1$ is independent of $p_1$ because $q_1$ is fixed only by the 
number of $p_2$-plaquettes. 
(d) $p_1$ vs $q_0$ for $p_2=0.01,0.03,0.05$ from above.
For very small $p_1$, $q_0$-sites are unnecessary because of sufficient
$q_1$-links supplied to compose $p_2$-plaquettes.
}
\end{figure}

\begin{figure}[h]
\includegraphics[width=8cm]
{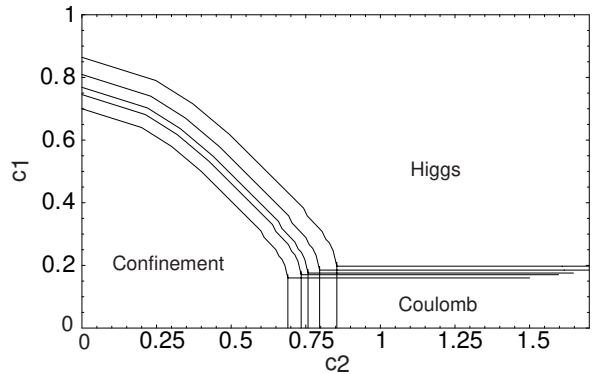}
\caption{\label{fig3}
The phase diagram in the $c_2-c_1$ plane determined
by the mean-field theory for $p = 0, 0.015, 0.022, 0.035, 0.05$
from below.  As the disorder increases, the region of
 Higgs phase diminishes.
}\end{figure}

\begin{figure}[htb]
\includegraphics[width=4cm]
{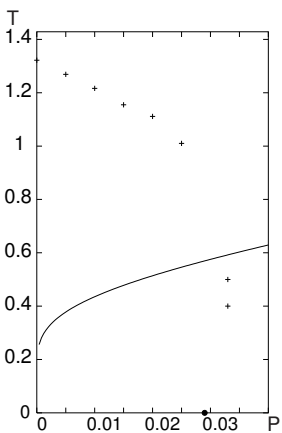}
\caption{\label{fig4}
Phase diagram in the $p-T$ ($p=p_2, T=c_2^{-1}$) plane 
for $c_1=0$ from Ref.\cite{ohno}. 
The phase boundary $p = p_c(T)$ 
gives the maximum value $p \simeq 0.0332\sim0.033$
at $T \simeq 0.3 \sim 0.4$. It ends 
at $T=0$ with $p \simeq 0.029$\cite{wang}. 
The solid curve is the Nishimori
line\cite{nishimori}  on which the ``thermal" effects valances with
the ``random" effects. 
}\end{figure}

\begin{figure}[h]
\includegraphics[width=6cm]{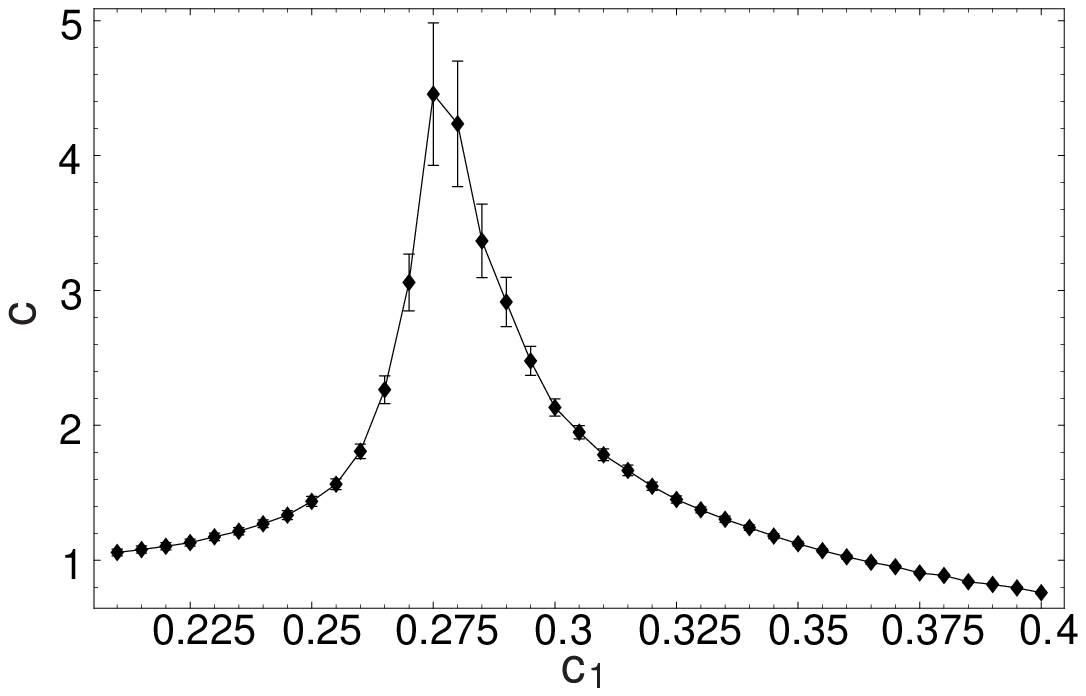}
\includegraphics[width=6cm]{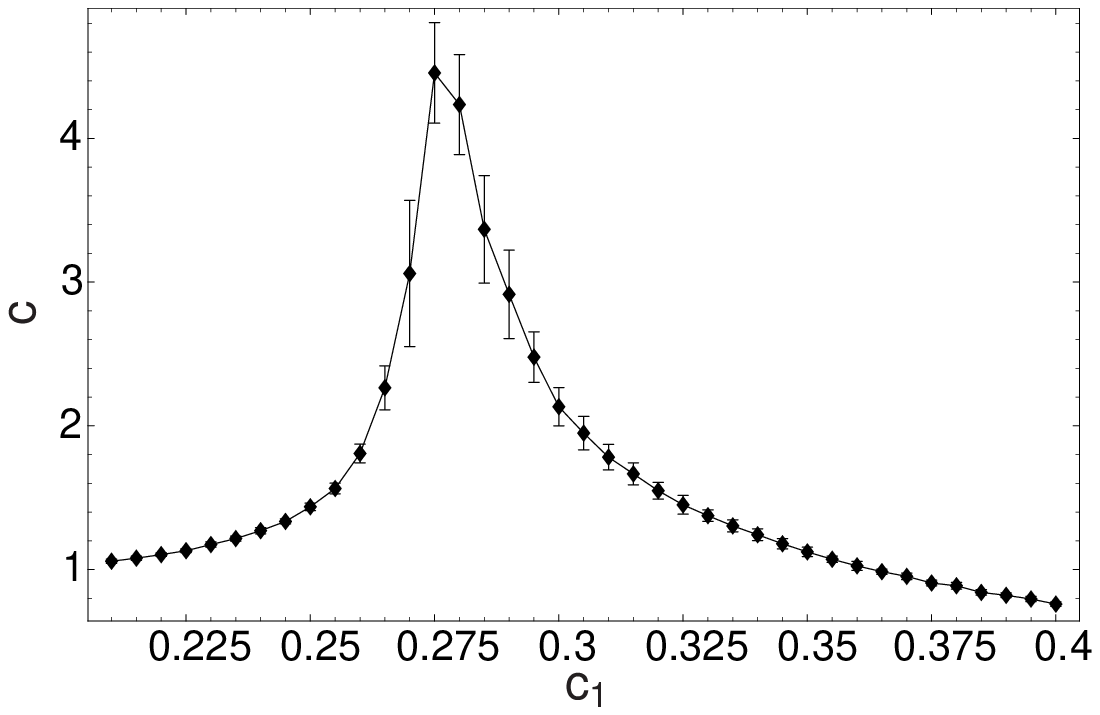}
\vspace{-0.2cm}\\
\hspace{3cm}(a) \hspace{5cm} (b) 
\caption{\label{fig5}
Specific heat for $c_2=1.0$ and $p=0.04$ with 
$L=24$, $N_s=40$.
(a) With MC errors and (b) With standard deviation over samples
(SDS).
}\end{figure}

\begin{figure}[h]
\includegraphics[width=7cm]{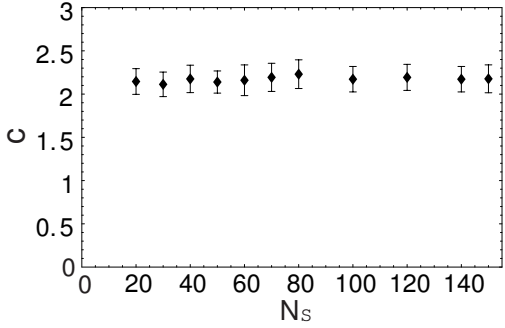}
\caption{\label{fig6}
Sample number ($N_s$) dependence of the specific heat
for $c_2=1.0, c_1=0.27$ with $p=0.03$
and $L=12$ with SDS.
}
\end{figure}

\begin{figure}[h]
\hspace{1cm}\includegraphics[width=8cm]{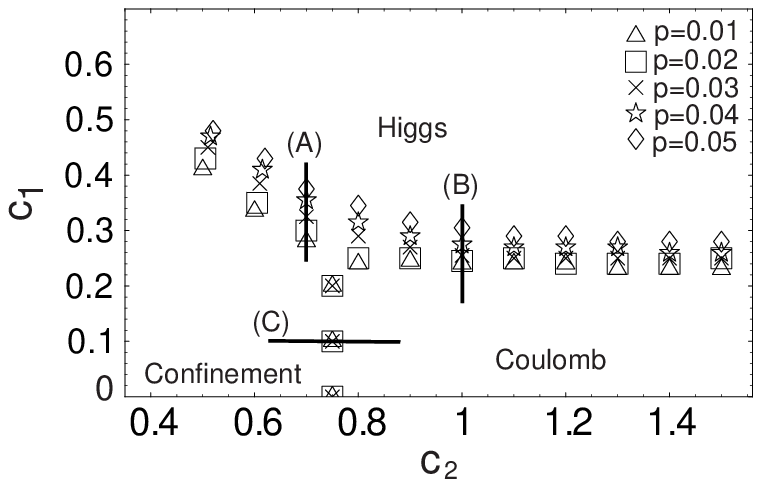}
\vspace{-0.2cm}\\
\hspace{5cm}(a)
\\
\\
\includegraphics[width=7cm]{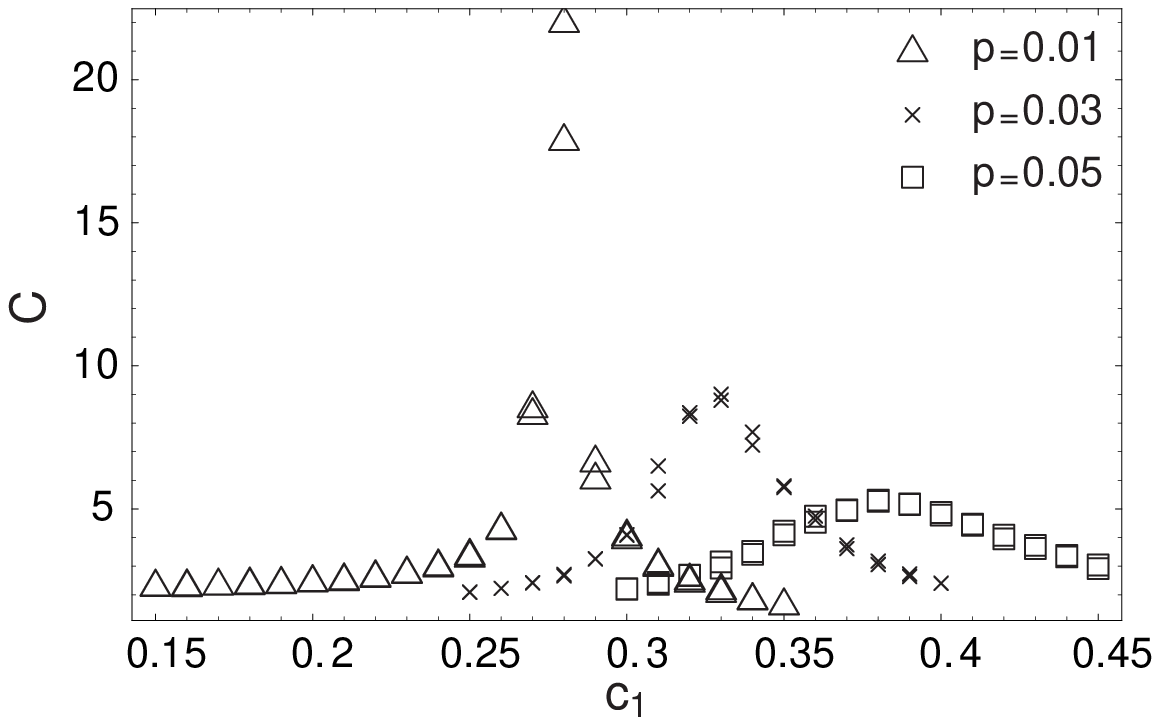}
\includegraphics[width=6.8cm]{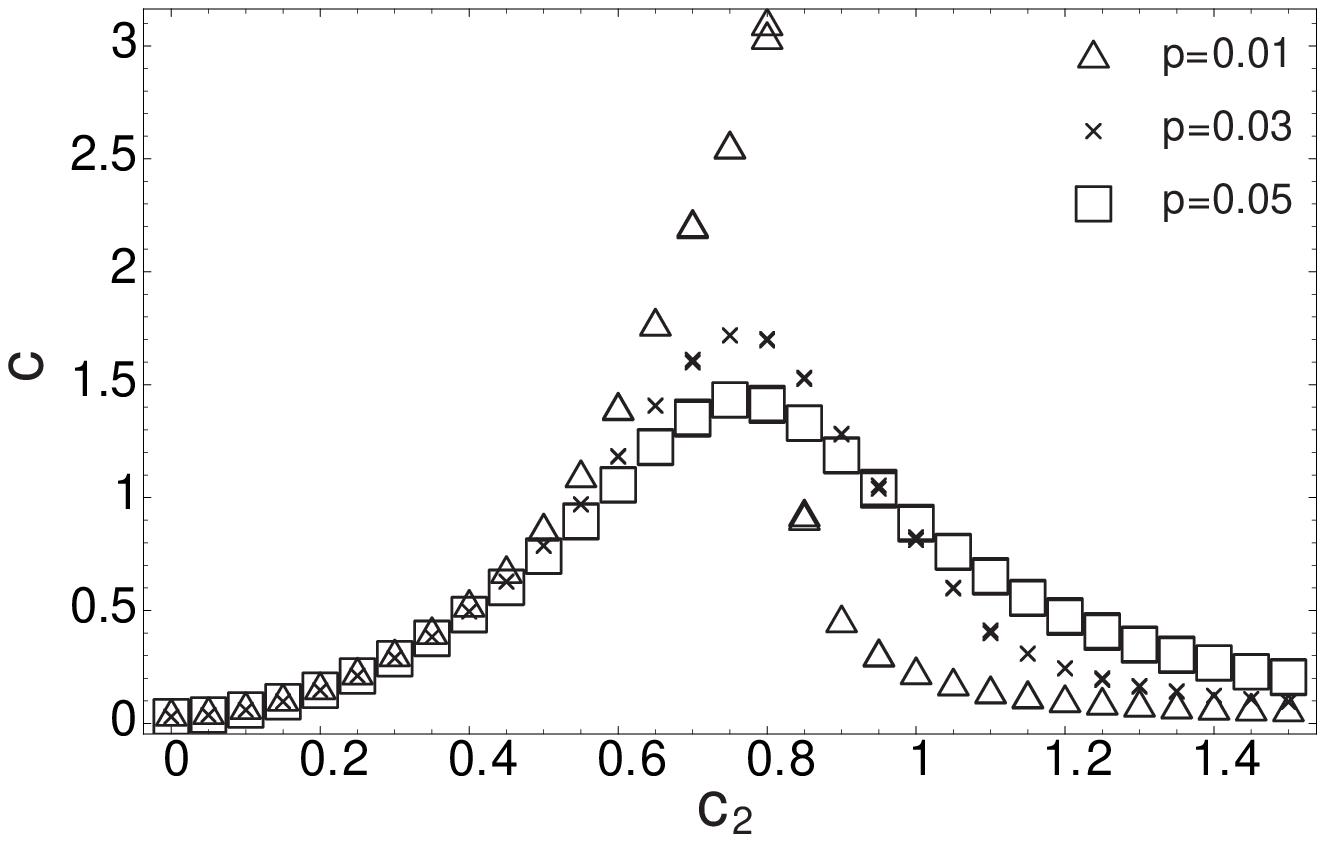}
\vspace{-0.2cm}\\
\hspace{3cm}(b) \hspace{5cm} (c) \\
\caption{\label{fig7} (a) Phase diagram in the $c_2-c_1$ plane with the
possible phase boundary curves determined by the location
of the peak of specific heat $C$ with $L=12$. The three segments
marked by (A,B,C) show the cases we shall examine in detail below.
(b) $p$-dependence of the specific heat  $C$ for $c_2=1.0$
with $L=12$. 
(c) $p$-dependence of the specific heat  $C$ for $c_1=0.1$
with $L=12$.}
\end{figure}

\begin{figure}[h]
\includegraphics[width=7cm]{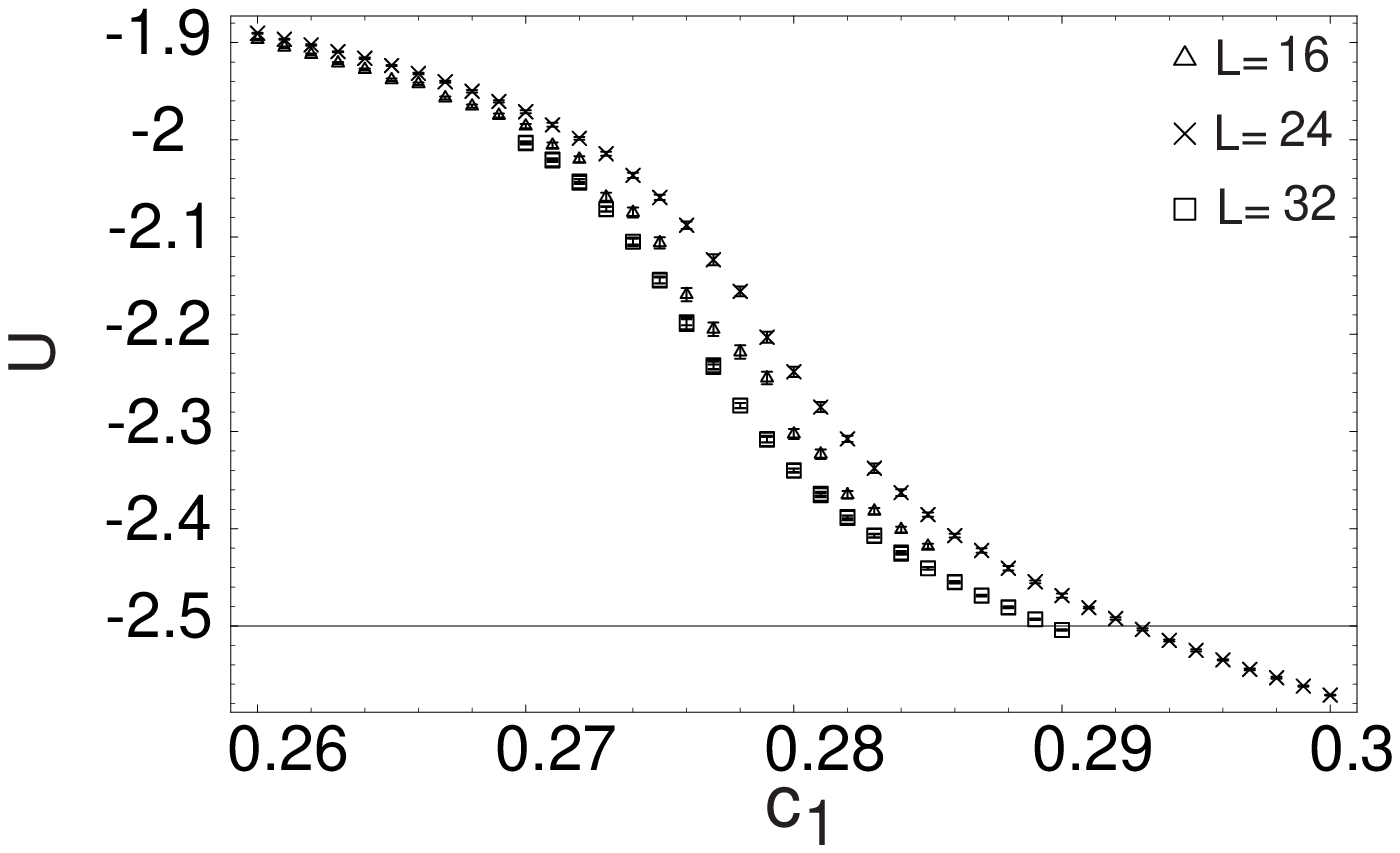}
\includegraphics[width=7cm]{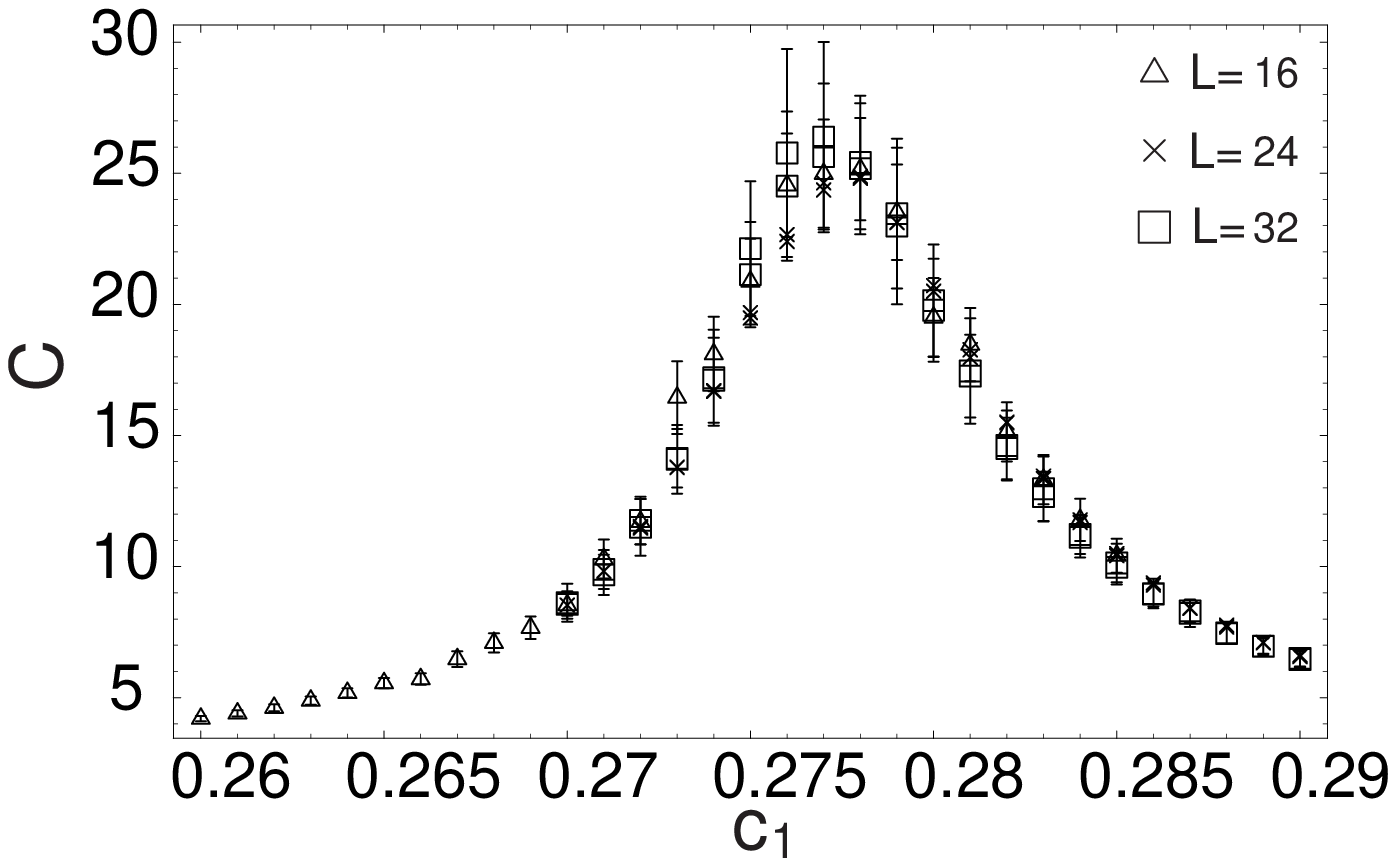}
\vspace{-0.2cm}\\
\hspace{4cm}(a) \hspace{6cm} (b) \\
\caption{\label{fig8}
Inernal energy $U$ and the specific heat $C$ for 
the case (A) $c_2=0.7$ at $p=0.01$.
(a) $U$ has no hysteresis and (b) $C$ has no systematic size dependence.
}\end{figure}

\begin{figure}[h]
\includegraphics[width=7cm]{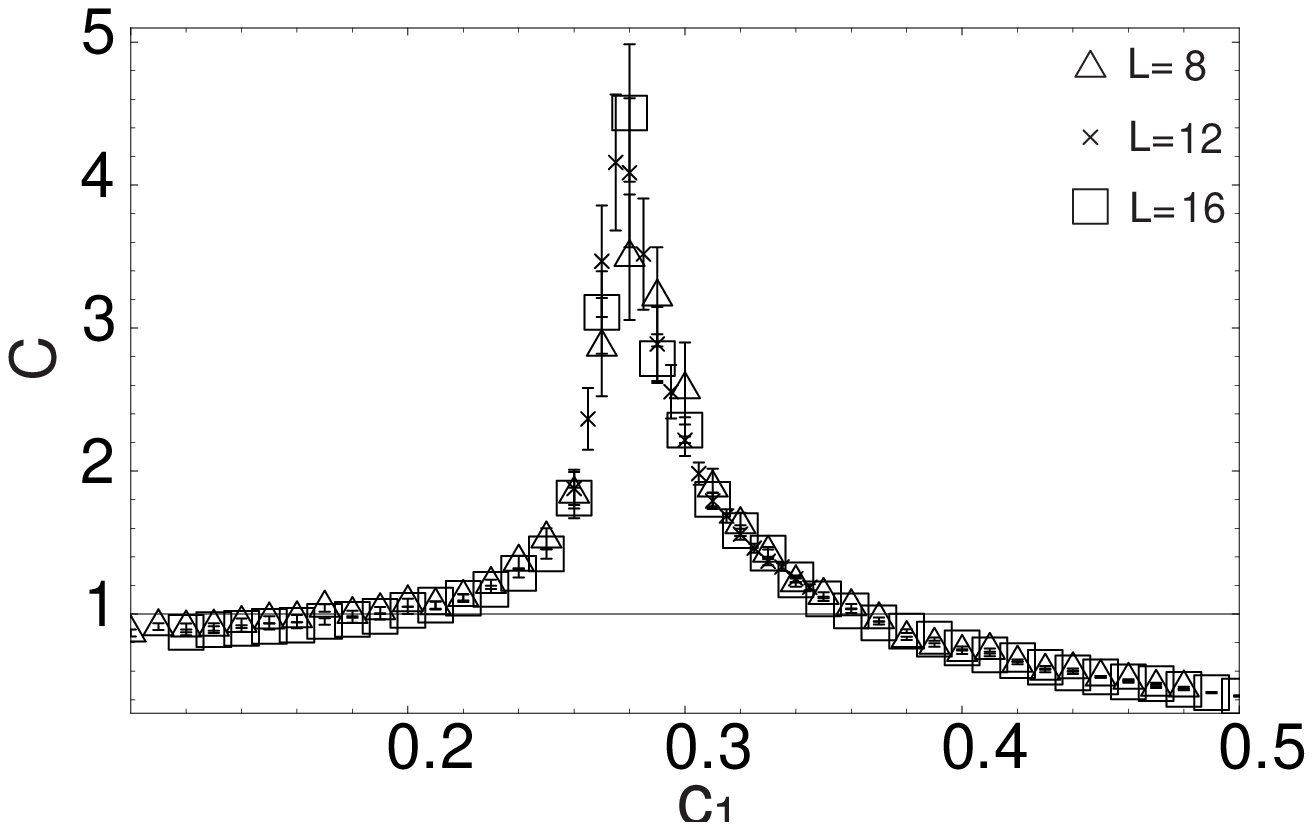}
\includegraphics[width=7cm]{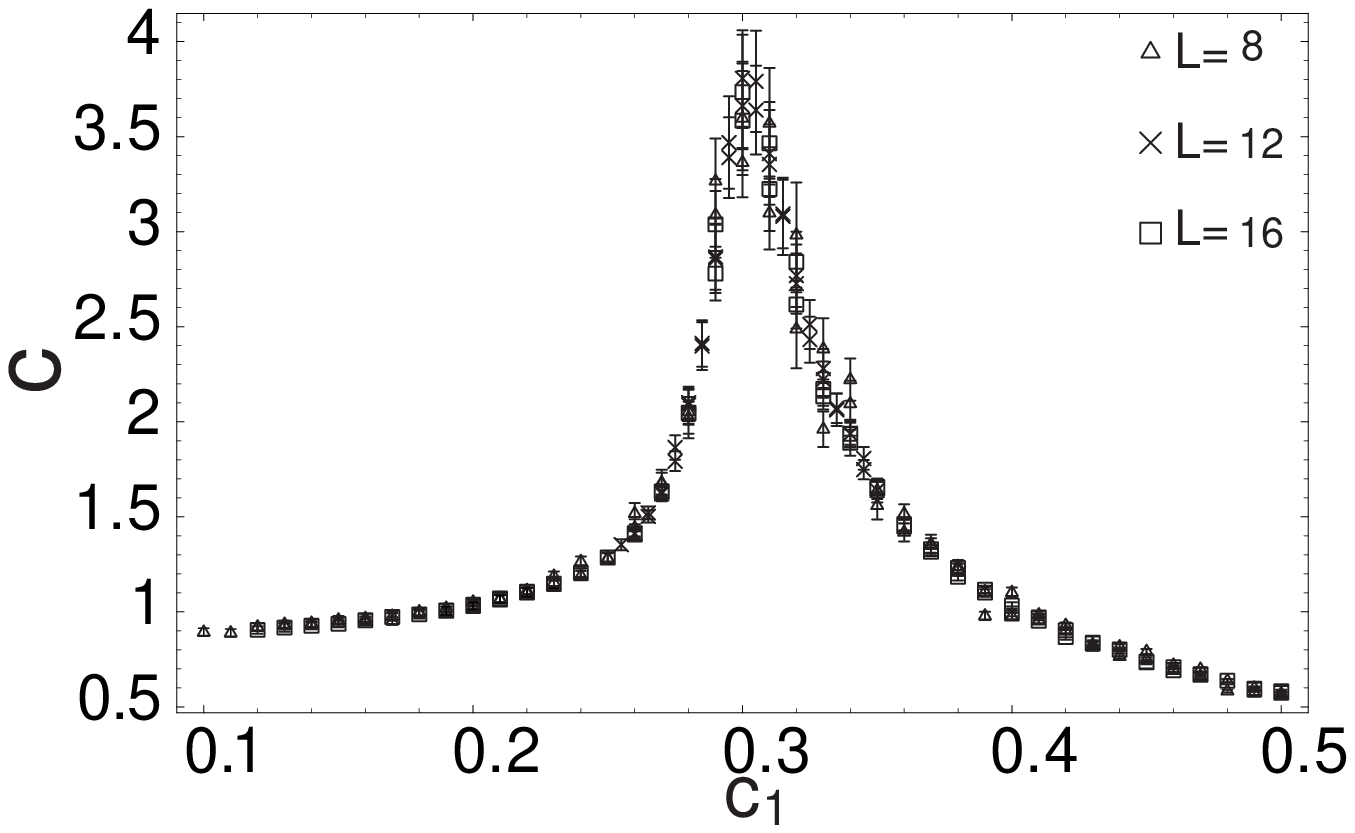}\vspace{-0.2cm}\\
\hspace{3cm}(a) \hspace{7cm} (b) \\
\\
\includegraphics[width=7cm]{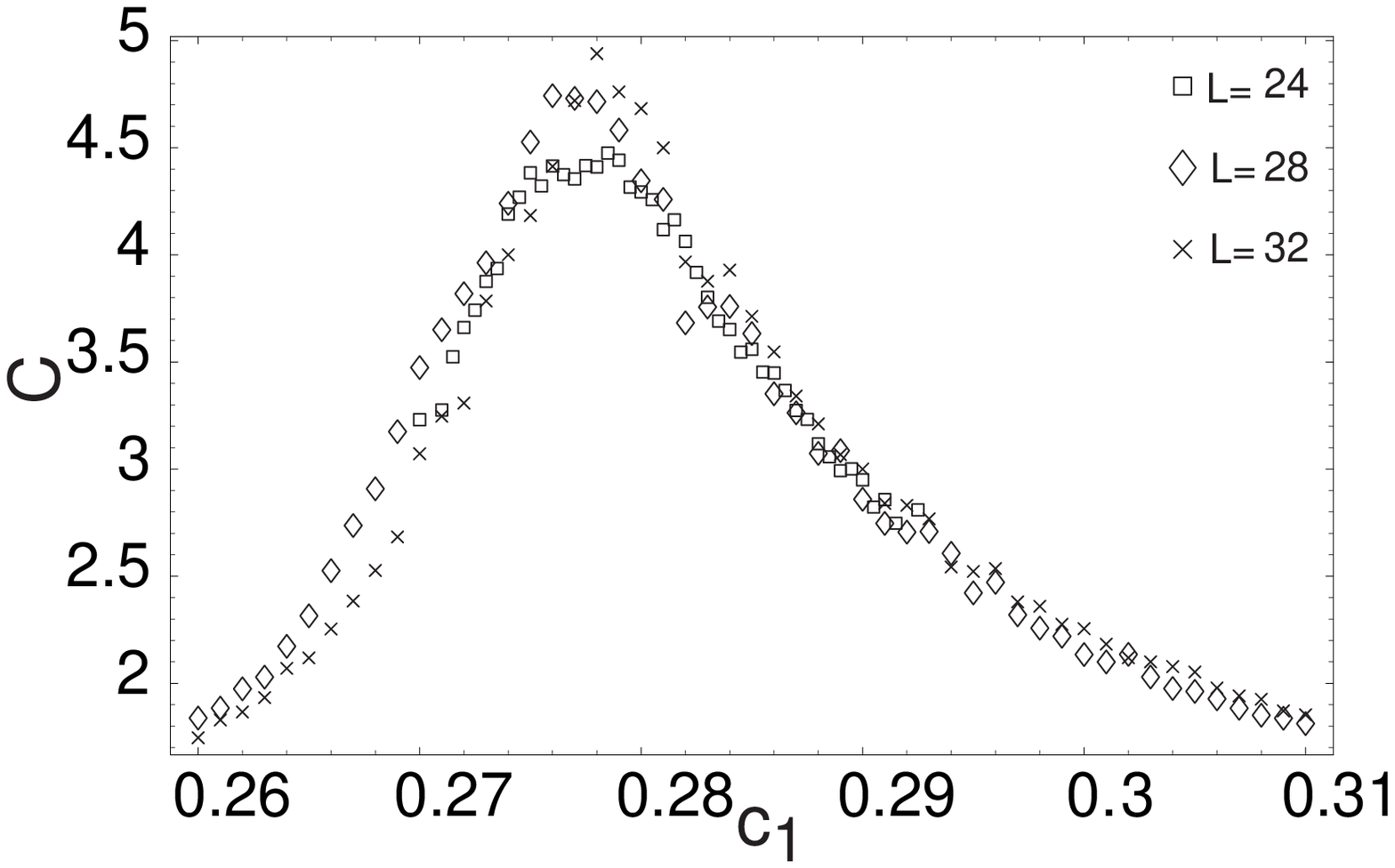}
\includegraphics[width=7cm]{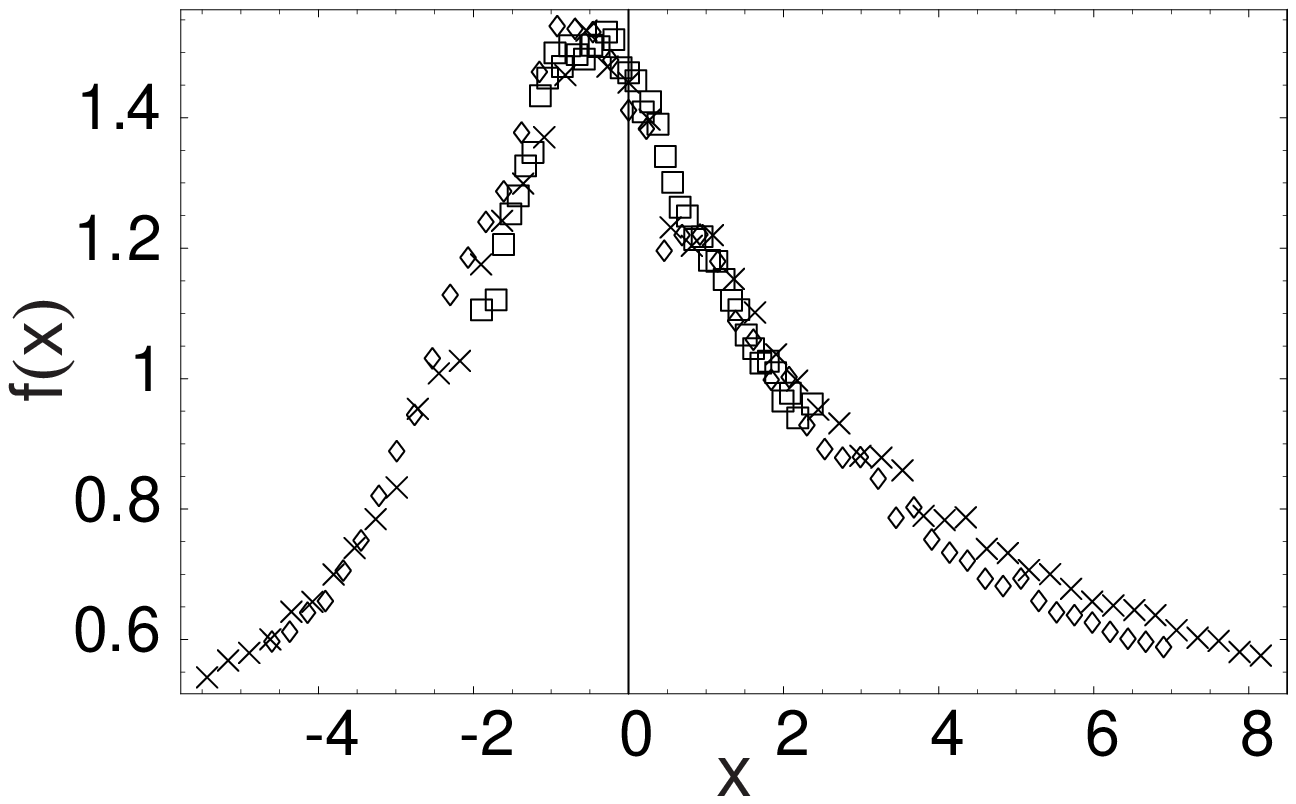}\vspace{-0.2cm}\\
\hspace{3cm}(c) \hspace{7cm} (d) \\
\caption{\label{fig9}
Specific heat $C$ for the case (B) $c_2=1.0$. 
(a) $C$  for $p=0.04$;
(b) $C$  for $p=0.05$.
(c) $C$  for $p=0.04$ before scaling;
(d) $C$  for $p=0.04$ after scaling.
}\end{figure}

(B) $c_2=1.0$:

For $p=0$ this case gives rise to a second-order
phase transition (See Fig.\ref{fig1}). 
For $p=0.01, 0.02, 0.03, 0.04$ the specific hear shows systematic
size-dependent development, which supports that the second-order transition
survives up to $p = 0.04$. The peak of $C$ for $p=0.05$ fails to show
systematic size dependent development.
In Fig.\ref{fig9}a,b, we compare $C$ for $p=0.04$ and $p=0.05$. 
Even for the size up to $L=16$, the difference appears clearly.
To confirm the existence of second-order transition for $p=0.04$,
we fit $C$ of $p=0.04$ with $L=24,28,32$ shown in 
Fig.\ref{fig9}a  by the finite-size  scaling 
hypothesis\cite{scaling} which reads for $C$ as
\be
C(c_1, L) = L^{\frac{\sigma}{\nu}}\ f(L^{\frac{1}{\nu}}\epsilon),
\ \ \ 
\epsilon \equiv \frac{c_1 -c_{1\infty}}{c_{1\infty}},
\label{scaling}
\ee
where $f(x)$ is the scaling function,
 $c_{1\infty}$ is the critical value of $c_1$ for the infinite system
$L = \infty$, and $\sigma,\ \nu$ are the critical exponents.
In Fig.\ref{fig9}b we present $f(x)$ calculated with the choice    
\be
c_{1\infty}= 0.277,\ \nu=0.90,\ \sigma= 0.273,
\ee
which supports the existence of $f(x)$.
For $p=0.05$ such a fit was impossible due 
to the lack of systematic size dependence.\\

\begin{figure}[h]
\includegraphics[width=7.5cm]{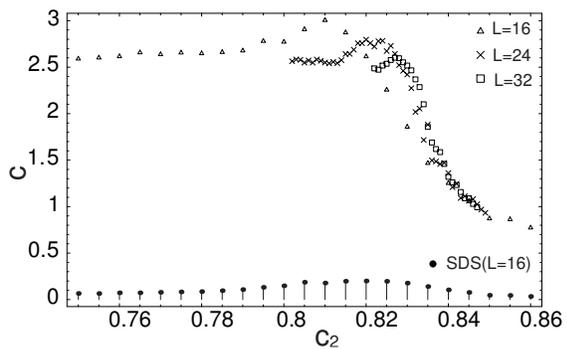}
\includegraphics[width=8cm]{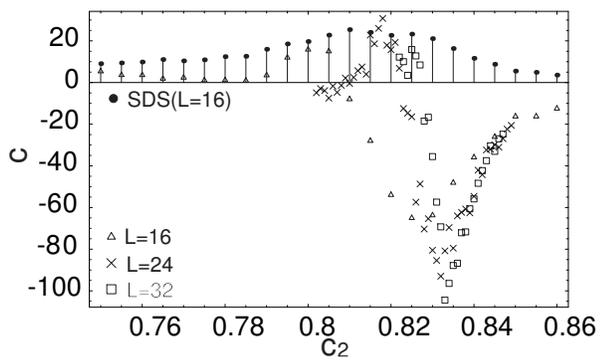}\vspace{-0.2cm}\\
\hspace{4cm}(a) \hspace{7cm} (b) \\
\\
\includegraphics[width=7.5cm]{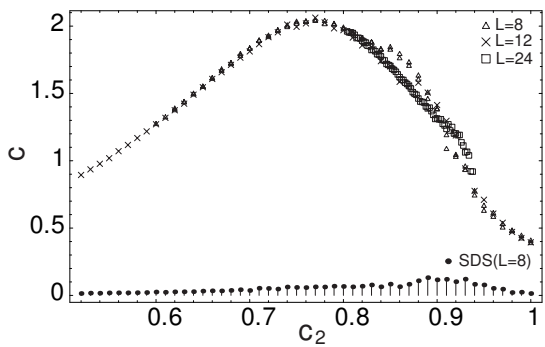}
\includegraphics[width=7.5cm]{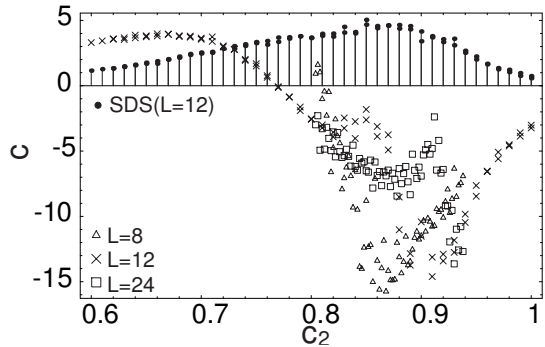}\vspace{-0.2cm}\\
\hspace{4cm}(c) \hspace{7cm} (d) \\
\caption{\label{fig10}
$C$ and  $dC/dT$ for the case 
(C) $c_1=0.1$. (a) $C$ for $p=0.01$, (b) $dC/dT$ for $p=0.01$,
(c) $C$ for $p=0.02$, (d) $dC/dT$ for $p=0.02$.
In each figure the SDS are shown separately.
}\end{figure}

(C) $c_1 =0.1$:

For $p=0$ we know that this case gives rise to a second-order
phase transition (See Fig.\ref{fig1}).
We shall argue that this transition changes to a third-order one
for  $p=0.01$ and becomes a  higher-order one or a cross over
for $p=0.02$.  
In Fig.\ref{fig10}a we present a close up of $C$ near $c_2 = 0.80$
for $p=0.01$.\footnote{By comparing 
Fig.\ref{fig10}a with $C$ for $c_1=0$\cite{ohno} 
we find the  two peaks of $C$
for $L=24$ locate at almost the same place $c_2 \simeq 0.82$.}
It shows that a small peak is developed, which shifts,
as $L$ increases, to larger $c_2$ direction
and diminishes gradually. 
We interprete this behavior of $C$
implies that the system is just in the transient
point from the second-order transition to a higher-order one
or to a crossover.
We remember that the possibility of similar change from
a second-order transition 
to a higher-order one has been pointed out 
in Ref.\cite{ohno} for $c_1=0$ as one lowers
the critical value of $c_2^{-1}$ (See Fig.\ref{fig4}). 
To study this point in detail, we measured a new observable,
the derivative $dC/dT$ of $C$.
To introduce the ``temperature" $T$ we  multiply the action
by a factor $\beta \equiv T^{-1}$ and making the derivative
of $C$ w.r.t. $T$,
\be
\frac{dC}{dT} = {\rm Sample\ average\ of\ }
\Big[-\beta^4 \big(\la A^3 \ra - \la A\ra^3 \big) +\beta^3
\big(\la A^2 \ra - \la A\ra^2 \big) \big(3\beta \la A \ra -2
\big)\Big]/V,
\ee
and set $\beta=1$ finally. 
In Fig.\ref{fig10}b we present $dC/dT$. 
Its signature changes at $c_2 \simeq 0.83 (L=32)$ where  the small peak
locates. This may be a precursor of a third-order transition.
It is natural that this small peak disappears at $L \rightarrow \infty$
and a sharp edge(cliff) remains, which implies 
a thrid-order transition characterized by
 a finite gap $\Delta(dC/dT) \neq 0$. 
In Figs.\ref{fig10}c,d we present $C$ and $dC/dT$ for $p=0.02$.
There are no indications of possible third-order transition.
They may describe a smoother transition or a crossover. 
The detailed study is necessary to calculate  $p_c$
of the present case, which may involve third-order or higher-order
phase transition.
As explained above, this is consistent with the result of Ref.\cite{ohno}.

\section{Conclusion and Discussion}
\setcounter{equation}{0} 

In this paper, we considered the 3D random Z(2) Higgs lattice
gauge theory  and studied
its global phase structure in the $c_2-c_1$ plane
by a new MFT and MC simulations.
The MC simulations showed that, for the case (A) of 
confinement-Higgs transition, 
the first-order transition for $p=0$ disappears quickly
at $p=0.01$.
For the case (B) of Coulomb-Higgs transition, 
the second-order transition for $p=0$ persists
up to $p=0.04$. The result for $p=0.05$ is failed to fit 
the scaling law.
For the case (C) of confinement-Coulomb transition, 
as $L$ increases, a small peak in $C$ developes for $p=0.01$ and 
then diminishes, which leads to
a third-order transition. However, for $p=0.02$, a third-order transition
has not been observed, so further study is necessary to explore
a possible higher-order transition.

Let us discuss some implications of these results for
a quantum memory. Inclusion of the Higgs coupling
produces the third Higgs phase which has the least  fluctuations of variables
among the three phases, and so more stable functions than in the Coulomb
phase are expected. 
The estimated critical concentration $p_c \simeq 0.04 \sim 0.05$ for the case (B)
is larger than $p_c \simeq 0.033$ for $c_1=0$. 
Of course, the nonlocal observables
become more complicated than the Wilson loop itself\cite{marc}.  
 
Concerning to the neural network, ``thermal" fluctuations 
previously considered as noise effects in Ref.\cite{kemukemu, nnet} 
act uniformly in space. 
In contrast, the random effects considered
in the present model are 
inhomogeneous.
However, these two effects seem to have no crucial qualitative differences
for a system staying near a phase boundary but just inside
 the Higgs phase. Either introduction of  tiny amount ofrandomnessin the 
 local coupling constants
or heating the sytem by tiny amount of temperature rise necessarily
drives the system into the Coulomb or confinement phase. 
  
When one tries to include the time evolution to a 3D neural net
in a manner faithful to quantum theory and/or statistical mechanics,
one faces a 4D system (with the imaginary-time as the fourth direction). 
Then the effect of randomness to
such a 4D system in of interest. Some analyses have appeared for
4D Z(2) RPGM\cite{arakawa}. Including the Higgs coupling
to this 4D model may be an interesting  subject to study.



\end{document}